\documentclass[final,twocolumns,5p]{elsarticle}
\usepackage{graphicx}
\usepackage{amssymb}
\usepackage{amsmath}
\usepackage{times}
\usepackage[outdir=./]{epstopdf}

\usepackage{xcolor} 

\makeatletter
\def\ps@pprintTitle{%
	\let\@oddhead\@empty
	\let\@evenhead\@empty
	\def\@oddfoot{\reset@font\hfil\thepage\hfil}
	\let\@evenfoot\@oddfoot
}
\makeatother

\begin{document}

\title{Analytical Phase Reduction for Weakly Nonlinear Oscillators}
\author{Iv\'an Le\'on}
\author{Hiroya Nakao}
\address{Department of Systems and Control Engineering, Tokyo Institute of Technology, Tokyo 152-8550, Japan}

\date{\today}

\begin{abstract}
%
%

Phase reduction is a dimensionality reduction scheme to describe the dynamics of nonlinear oscillators with a single phase variable. While it is crucial in synchronization analysis of coupled oscillators, analytical results are limited to few systems. In this work, we analytically perform phase reduction for a wide class of oscillators by extending the Poincar\'e-Lindstedt perturbation theory. We exemplify the utility of our approach by analyzing an ensemble of Van der Pol oscillators, where the derived phase model provides analytical predictions of their collective synchronization dynamics. 

\end{abstract}

  \maketitle
  
\section{Introduction}
Populations of coupled oscillators are at the core of many natural and technological systems, from the beating of the heart to the power transmission in the electrical grid \cite{PRK01}. Although the relevance of this topic has motivated extensive research in the last century, the large number of degrees of freedom and intrinsic complexity of those systems have hampered complete comprehension.

In the '60s and '70s, the seminal works of Winfree \cite{Win67,Win80} and Kuramoto \cite{Kur75,Kur84} boosted our understanding of oscillatory dynamics by introducing phase reduction. Phase reduction is a dimensionality reduction scheme that, assuming weak coupling, derives a model capturing the dynamics of each oscillator in terms of a single variable: the oscillator's phase $\theta$. Derivation of the Kuramoto model, which provided an analytical explanation of the synchronization transition \cite{Kur75,Str00,PRK01}, clearly demonstrated the advantages of phase reduction.

Since then, there has been an explosion of research employing phase models to describe numerous systems, such as ensembles of neurons \cite{BGL+20}, mechanical oscillators \cite{mertens11}, or flashing of fireflies \cite{EK86}, to cite a few. In addition, there have also been multiple advances in phase reduction, as generalizations to include noise \cite{Gol10,Jun09}, non-weak coupling \cite{leon20,Kure22}, higher-order corrections \cite{leon19,WilErm2019,KuraNakao19,RP19}, or to deal with slow-fast systems \cite{ZKN22}. Despite these advances and the usefulness of phase reduction, analytical results are limited only to a few types of oscillators for which the limit cycle and isochrons are easily computable \cite{Win80,Win79}.

In this work, we present an analytical perturbative approach to phase reduction for oscillators whose unperturbed solution is known. The paradigmatic example is weakly nonlinear oscillators that reduce to the harmonic oscillator in the absence of nonlinearities. We achieve the phase reduction by computing the frequency, limit cycle, and phase sensitivity function (a.k.a. infinitesimal phase response curve) as a perturbative expansion around the known solution. 
We illustrate the usefulness of the approach by analyzing an ensemble of globally coupled Van der Pol oscillators through the reduced phase model, obtaining analytical predictions of the collective 
dynamics.

The method presented in this work broadens our knowledge about analytical phase reduction and provides a useful tool to better understand the oscillatory dynamics for a wide variety of systems. It derives simple, analytical, and analyzable phase models that capture the weak coupling dynamics, as exemplified by the results obtained for the Van der Pol oscillator, a well-known model with many experimental applications \cite{PRK01,GM88}.
%
%

\section{Phase reduction}
Before presenting our results, it is convenient to make a brief introduction to phase reduction \cite{Nak16}. Consider an $m$ dimensional oscillator whose state is defined by $\boldsymbol{X}=(X_1,\dots,X_m)$. The dynamics of the oscillator under the effect of some small perturbation $\epsilon \boldsymbol{p}$ is described by the ordinary differential equation (ODE):
	\begin{equation}\label{eq.generalosc}
		\dot{\boldsymbol{X}}=\boldsymbol{F}(\boldsymbol{X})+\epsilon \boldsymbol{p},
	\end{equation}  
where $\boldsymbol{F}(\boldsymbol{X})$ is the velocity field and overdot is the derivative with respect to time $t$. In the absence of perturbation, Eq.~\eqref{eq.generalosc} displays a linearly stable limit cycle $\boldsymbol{X}_c$ with frequency $\omega$. The perturbation $\boldsymbol{p}$ may depend on time $t$ and oscillator state $\boldsymbol{X}$. 

When the system is unperturbed, we define a phase variable $\theta$ on the limit cycle that grows constantly from $0$ to $2\pi$ in one period of the oscillation. This definition can be extended to the whole basin of attraction of the limit cycle using the concept of isochron \cite{Win80}. Isochrons foliate the state space and assign to each point their asymptotic phase, $\theta=h(\boldsymbol{X})$. Since the asymptotic phase of every point has to be computed to obtain the isochrons, an analytical expression for the isochrons is not generally accessible, except for a few types of oscillators with polar symmetry.

The evolution of the phase in the presence of a perturbation is computed from Eq.~\eqref{eq.generalosc} and the definition of phase. Assuming the perturbation to be weak compared to the second Floquet exponent $\lambda$, which characterizes the stability of the limit cycle, i.e., $\epsilon \ll \lambda/\omega$, we obtain a simple expression:
	\begin{equation}\label{eq.phasered}
		\dot{\theta}=\omega+\epsilon \boldsymbol{Z}(\theta) \cdot \boldsymbol{p},
	\end{equation}
where terms of order $\epsilon^2$ have been neglected~\cite{leon19,gengel21,leon22a} and $\boldsymbol{Z}(\theta)
=\nabla_{\boldsymbol{X}}\theta=\nabla_{\boldsymbol{X}}h(\boldsymbol{X}_c(\theta))$ is the gradient of the asymptotic phase $h({\boldsymbol X})$ evaluated on the limit cycle $\boldsymbol{X_c}$ at phase $\theta$. The function $\boldsymbol{Z}(\theta)$ is called the phase sensitivity function and characterizes how much the phase is shifted under the effect of an infinitesimal perturbation. In order for Eq.~\eqref{eq.phasered} to be closed, we approximately evaluate $\boldsymbol{p}$ on the limit cycle at $\boldsymbol{X_c}(\theta)$.
We note the big dimensionality reduction from the $m$-dimensional system Eq.~\eqref{eq.generalosc} to the one-dimensional oscillator Eq.~\eqref{eq.phasered} that captures the dynamics for small $\epsilon$.

From this derivation, we learned that to perform phase reduction, we only need expressions for the frequency $\omega$, the limit cycle $\boldsymbol{X_c}(\theta)$, and the phase sensitivity function $\boldsymbol{Z}(\theta)$. Although 
$\boldsymbol{Z}(\theta)$ can be directly computed from the asymptotic phase, $h(\boldsymbol{X})$, it is easier to employ the Floquet theory. This theory states that the phase sensitivity function is the periodic solution to the adjoint equation:
	\begin{equation}\label{eq.adj}
		\omega\frac{d \boldsymbol{Z}(\theta)}{d\theta}=-J^\top(\boldsymbol{X}_c(\theta)) \boldsymbol{Z}(\theta)
	\end{equation}
subject to the normalization condition $\boldsymbol{Z} (\theta) \cdot\boldsymbol{F}({\boldsymbol X}_c(\theta)) =\omega$ \cite{HI97,Izh07,ermentrout96,BMH04}, where $J^\top(\boldsymbol{X}_c(\theta))$ is the transpose of the Jacobian matrix of $\boldsymbol{F}$ evaluated on the limit cycle at ${\boldsymbol X}_c(\theta)$. 

\section{Perturbative method}

In this work, we perform analytical phase reduction for a general class of oscillators whose evolution depends on a parameter $\mu$:
	\begin{equation}\label{eq.pertosc}
		\dot{\boldsymbol{X}}=\boldsymbol{F}(\boldsymbol{X},\mu),
	\end{equation}	
which reduces to an oscillator whose solution $\boldsymbol{X}_0$ for $\mu=0$ is known. Weakly nonlinear oscillators belong to this class since they reduce to a harmonic oscillator when nonlinearities are eliminated.

Provided that the solution for $\mu=0$ is known, we can develop a perturbation theory around this solution. One has to be careful at this step since ordinary perturbation theory produces secular terms, and thus, divergent solutions. We solve this issue by using the Poincar\'e-Lindstedt method, although other techniques such as multiple-timescale perturbation theory could also be applied \cite{Nayf93,Strogatz,ABF18}. We chose the Poincar\'e-Lindstedt method for its simplicity.

The first step in the Poincar\'e-Lindstedt method is to perform a change of variable from time to phase, $\theta=\omega t$, where $\omega$ is yet unknown. Then, we expand in powers of $\mu$ the frequency $\omega=\omega_0+\mu\omega_1+\dots$ and the limit cycle solution $\boldsymbol{X}=\boldsymbol{X}_0+\mu \boldsymbol{X}_1+\dots$, and replace them in \eqref{eq.pertosc}. 
	\begin{equation}
	\omega_0 \frac{d\boldsymbol{X}_0}{d\theta}+\mu\left(\omega_1 \frac{d\boldsymbol{X}_0}{d\theta}+\omega_0 \frac{d\boldsymbol{X}_1}{d\theta}\right)+\dots=\boldsymbol{F}(\boldsymbol{X}_0+\mu \boldsymbol{X}_1+\dots,\mu)
	\end{equation}	
Expanding the right-hand side and collecting the terms at each order of $\mu$, an ODE for each order is obtained.

We solve the ODEs recursively subject to an initial condition ${\boldsymbol X}(0) = {\boldsymbol A}_0 + \mu {\boldsymbol A}_1 + \mu^2 {\boldsymbol A}_2 \cdots$ where ${\boldsymbol A}_n$ will be determined later.
The solutions to the ODEs, as in ordinary perturbative theory, contain secular terms. However, now these terms depend on the undetermined constants $\boldsymbol{A}_n$ and unknown frequencies $\omega_{n}$. We can uniquely choose those quantities so that secular terms are removed. This method yields the frequency $\omega$ and the limit cycle solution expressed in terms of the phase $\theta$ to any desired order $\mu$.

\begin{figure}
	\includegraphics[width=\linewidth]{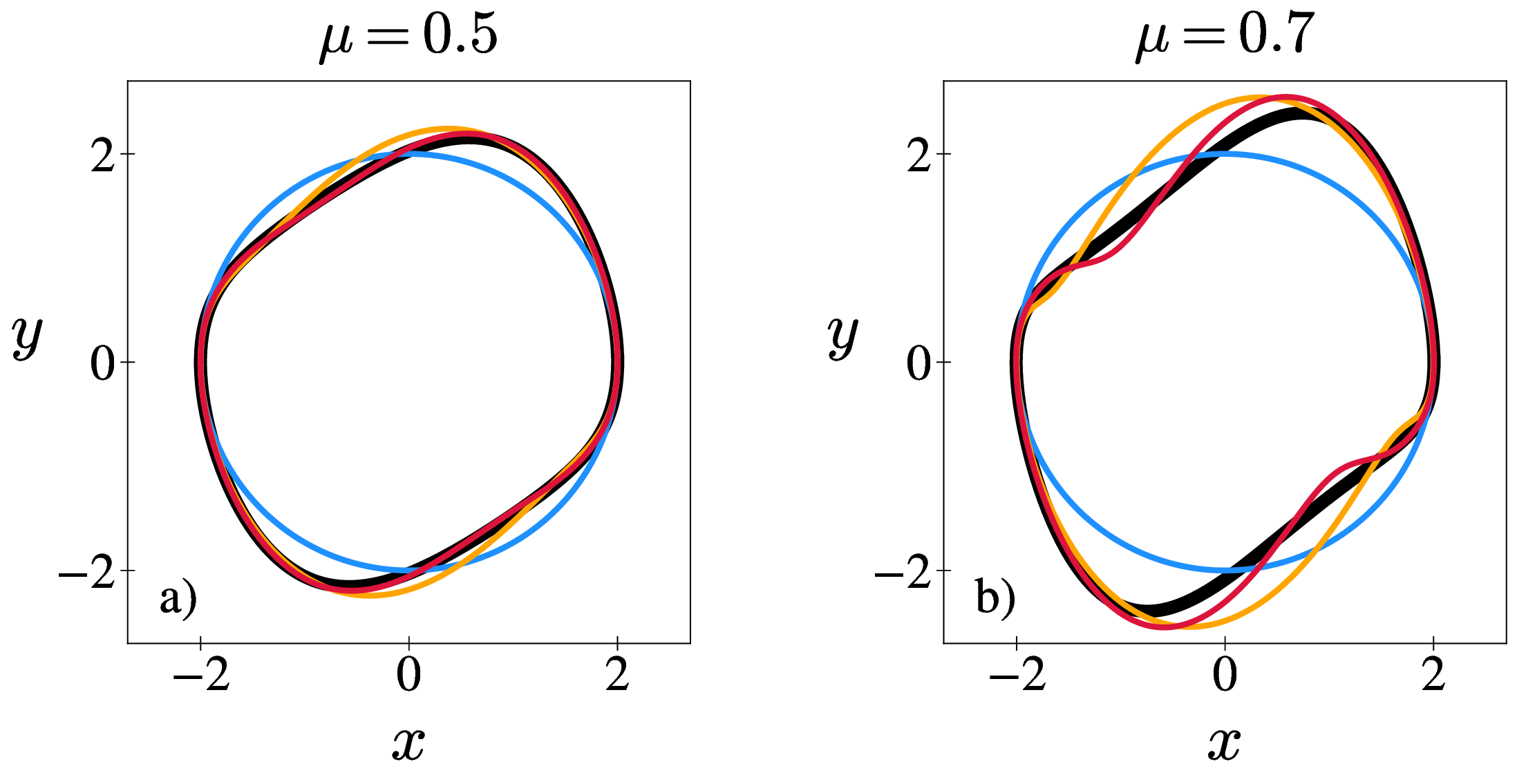}
	\caption
	{Limit cycle of the Van der Pol oscillator for $\mu=0.5$ (a) and $\mu=0.7$ (b). The exact numerical limit cycle and the $O(1)$, $O(\mu)$ and $O(\mu^2)$ approximations are depicted in black, blue, orange, and red, respectively.}
	\label{fig.LC}
\end{figure}

As an example, we apply the method to the Van der Pol oscillator \cite{VdP26,VdP27}:
\begin{eqnarray}
	\dot{x}&=&y,\nonumber\\
	\dot{y}&=&-x+\mu(1-x^2)y.\label{eq.vandp}
\end{eqnarray}
The analytic expressions obtained with the above procedure and their explicit derivation can be found in \ref{app.VP}. In Fig.~\ref{fig.LC}, we compare the approximate limit cycles with the exact ones obtained by direct numerical simulation for $\mu=0.5$ (a) and $\mu=0.7$ (b). First of all, we remark that the exact limit cycle (black) is far from the perfect circle of the unperturbed harmonic oscillator (blue), i.e., the $O(1)$ approximation. The higher-order corrections contain new harmonics that reduce this discrepancy, as can be seen in the $O(\mu)$ (orange) and $O(\mu^2)$ (red) approximations. As expected, the accuracy improves as higher orders are considered, while decreases as $\mu$ increases. Strikingly, the order $\mu^2$ approximation is accurate even for non-small values of $\mu$, as can be seen in panel~(b).

The crucial idea in this work is the extension of the Poincar\'e-Lindstedt method to compute the phase sensitivity function $\boldsymbol{Z}$ by applying a similar perturbative scheme to the adjoint equation \eqref{eq.adj}. We expand the phase sensitivity function in powers of $\mu$: $\boldsymbol{Z}=\boldsymbol{Z}_0+\mu \boldsymbol{Z}_1+\dots$. Since $\boldsymbol{Z}(\theta)$ is a $2\pi$-periodic function, all elements in the expansion are also $2\pi$-periodic and do not contain secular terms. 
Plugging the expansion of $\boldsymbol{Z}$ and the expressions for the frequency and limit cycle into Eq.~\eqref{eq.adj}, we obtain
	\begin{align}
&	\omega_0 \frac{d\boldsymbol{Z}_0}{d\theta}+\mu\left[\omega_1 \frac{d\boldsymbol{Z}_0}{d\theta}+\omega_0 \frac{d\boldsymbol{Z}_1}{d\theta}\right]+\dots\nonumber\\
&	= -J^\top(\boldsymbol{X}_0+\mu \boldsymbol{X}_1+\dots)(\boldsymbol{Z}_0+\mu \boldsymbol{Z}_1+\dots)
	\end{align}
and the normalization condition becomes
	\begin{equation}
	\omega_0+\mu\omega_1+\dots=(\boldsymbol{Z}_0+\mu \boldsymbol{Z}_1+\dots)\cdot\boldsymbol{F}(\boldsymbol{X}_0+\mu \boldsymbol{X}_1+\dots,\mu).
	\end{equation}
Expanding the right-hand side of both equations and collecting the terms at each order of $\mu$, we obtain an ODE and normalization condition for each order. The solutions to the system contain secular terms, which should be removed by choosing appropriately the integration constants as in the standard Poincar\'e-Lindstedt method. This procedure yields the phase sensitivity function as a power expansion in $\mu$. See \ref{app.VP} for the derivation and analytical results on the Van der Pol oscillator.

\begin{figure}
	\includegraphics[width=\linewidth]{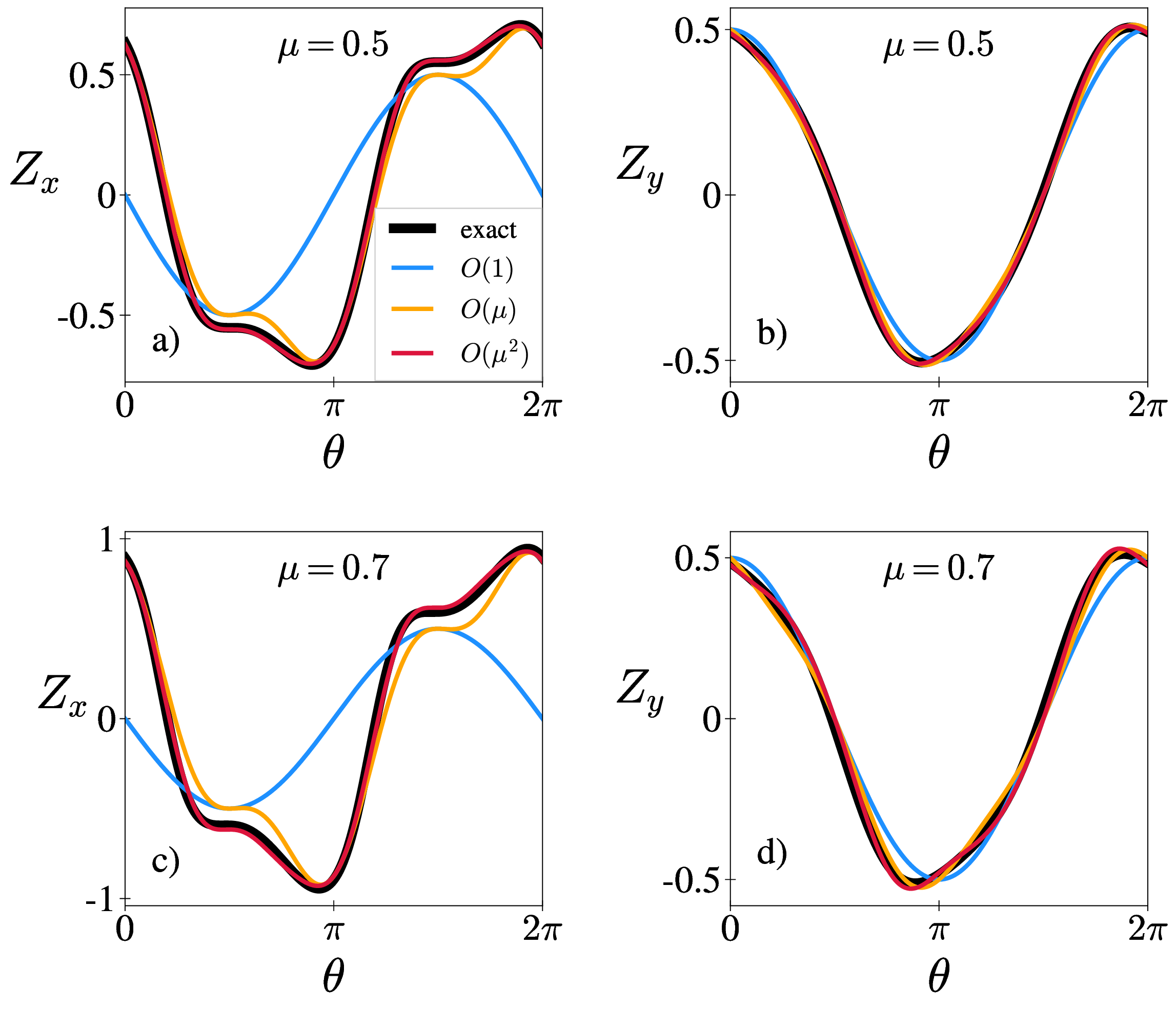}
	\caption
	{Phase sensitivity function ${\boldsymbol Z} = (Z_x, Z_y)$ ($x$ and $y$ components) of the Van der Pol oscillator for $\mu=0.5$ [(a) and (b)] and $\mu=0.7$ [(c) and (d)]. The exact numerical $\boldsymbol{Z}$ and the $O(1)$, $O(\mu)$, and $O(\mu^2)$ approximations are depicted in black, blue, orange, and red, respectively. }
	\label{fig.phasesensitivity}
\end{figure}

We now check the validity and accuracy of the method by computing the phase sensitivity function of the Van der Pol oscillator. We depict the $x$ and  $y$ components of $\boldsymbol{Z}$ vs the phase $\theta$ (see~\ref{app.VP} for the definition) in Figs.~\ref{fig.phasesensitivity} (a) and (b), respectively, for $\mu=0.5$. The exact results in black obtained by numerically solving the adjoint equation \eqref{eq.adj} are compared with the $O(1)$, $O(\mu)$, and $O(\mu^2)$ approximations in blue, orange, and red, respectively. Figures~\ref{fig.phasesensitivity} (c) and (d) depict the same components of $\boldsymbol{Z}$ for $\mu=0.7$. The conclusions about accuracy are analogous to those for the limit cycle. Remarkably, the $O(\mu^2)$ approximation is extremely accurate even for $\mu=0.7$, indicating the validity of the approximation up to those values. 

We note that, although the Poincar\'e-Lindstedt method is usually applied to weakly nonlinear oscillators, the above procedure is not limited to them. For example, it can be applied to oscillators that already possess a limit cycle at $\mu=0$, as shown for a Stuart-Landau oscillator with a constant bias in 
\ref{app.SLcb}.
 
Finally, we remark that being able to obtain analytical expressions for the frequency, limit cycle, and phase sensitivity function implies that we can 
analytically perform phase reduction. Moreover, since those quantities are independent of the perturbation ${\boldsymbol p}$, once they have been computed, they can be used irrespective of the coupling types or topology. Thus, it provides simple analytical phase models capturing the weak coupling dynamics.

\section{Ensemble of globally coupled Van der Pol oscillators}

We now illustrate how the reduced phase model obtained by the presented approach can help understanding the collective synchronization dynamics of oscillatory systems. We choose, as a simple example, a population of $N$ globally coupled Van der Pol oscillators with identical properties. The $i$th oscillator ($i=1, ..., N$) is coupled to other oscillators through:
	\begin{equation}\label{eq.coupling}
	\boldsymbol{p}_i(\boldsymbol{X}_1,\dots,\boldsymbol{X}_N)=\frac{\epsilon}{N}\sum_{k=1}^{N}\bigg((x_k-x_i)\cos\alpha ,(y_k-y_i)\sin\alpha \bigg),
	\end{equation} 
where $\epsilon$ is the coupling strength and $\alpha$ measures the ratio of $x$ and $y$ couplings. We remark that the parameter $\alpha$ is not 
the phase-lag of the Kuramoto-Sakaguchi model, as will be observed when phase reduction is performed. Denoting the phase of the $i$th oscillator as $\theta_i$ and replacing \eqref{eq.coupling} evaluated on the limit cycle, i.e., with ${\boldsymbol X}_i = {\boldsymbol X}_c(\theta_i)$, and the phase sensitivity function into \eqref{eq.phasered}, we analytically obtain the reduced phase model up to order $\mu^2$. We then perform averaging \cite{Kur84} to remove the fast-oscillating terms and obtain the phase model:
	\begin{multline}\label{eq.vnpglophase}
	\dot{\theta}_i=\Omega+\epsilon\bigg[\eta_{1s} R_1\sin (\Psi_1 -\theta_i)+\eta_{1c} R_1\cos (\Psi_1 -\theta_i)\\
	+\eta_3 R_3\sin (\Psi_3 -3\theta_i )\bigg],
	\end{multline}
where we have defined the Kuramoto-Daido order parameters $R_ke^{i\Psi_k}=\sum_j e^{ik\theta_j}$ for $k=1, 3$~\cite{Dai96}. The new constants are analytically represented by the original parameters as $\Omega=-1+\mu^2/16-\epsilon\eta_{1c}$, $\eta_{1c}=\mu (\sin\alpha+9\cos\alpha)/8$, $\eta_{1s}=\frac{1}{2} \left[\sin\alpha\left(1+\mu ^2/64\right) +\cos\alpha  \left(1-7 \mu ^2/64\right)\right]$, and $\eta_3=\mu^2\left(5\cos\alpha-3\sin\alpha\right)/128$. In contrast to the ensemble of the original Van der Pol oscillators, the reduced phase model \eqref{eq.vnpglophase} is straightforward to analyze following the standard techniques; see \ref{app.analy} for details. In what follows, we consider the thermodynamic limit $N\rightarrow\infty$.
 
\begin{figure}
	\includegraphics[width=\linewidth]{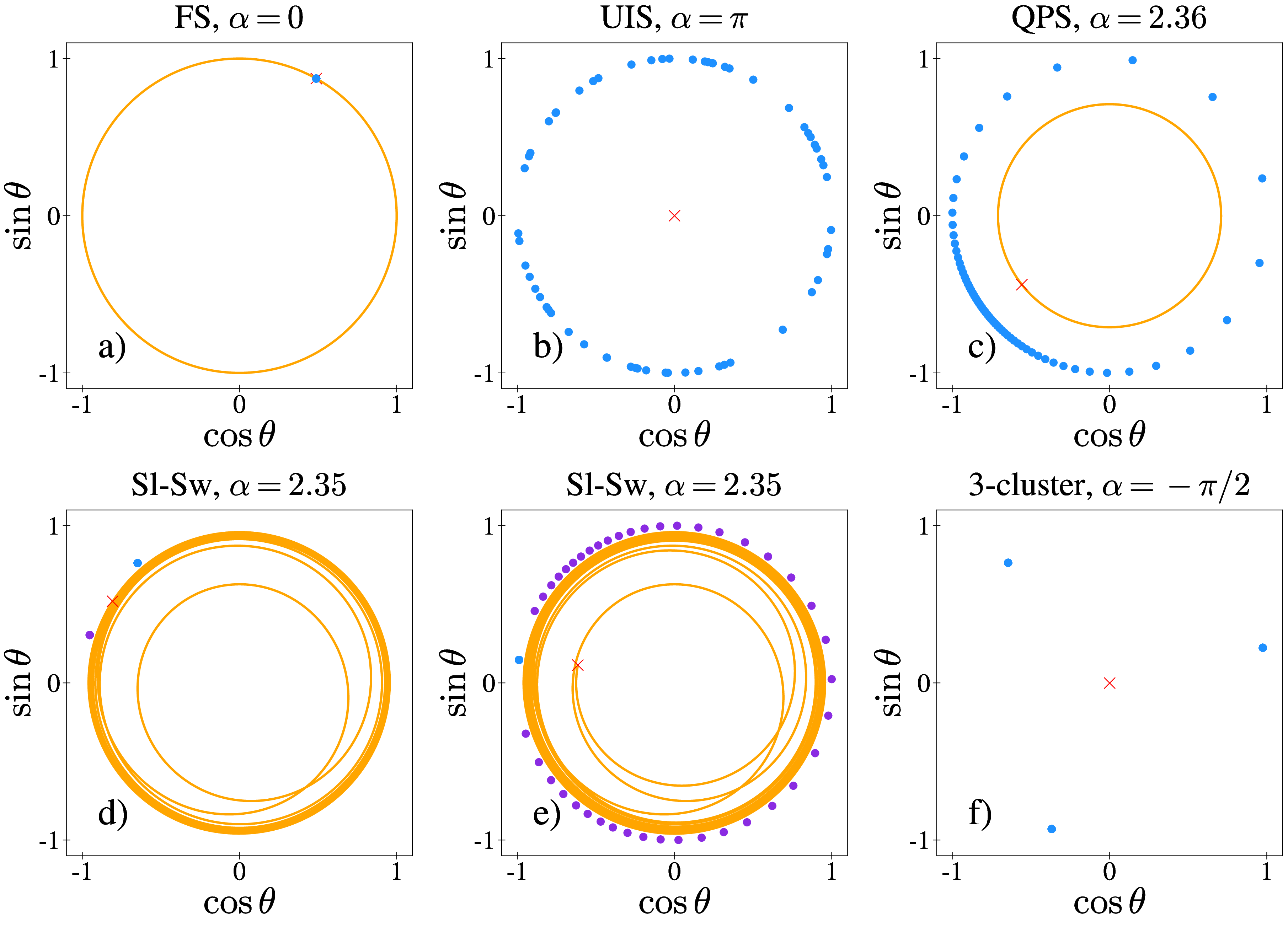}
	\caption
	{Snapshots of FS (a), UIS (b), QPS (c), slow switching (d,e), and three-cluster state (f) for $\epsilon=0.1$, $\mu=0.5$, and the displayed $\alpha$. 
	 The blue dots are 60 randomly chosen oscillators from a population of $N=300$ phase oscillators, the orange line is the evolution of the Kuramoto order parameter $(R_1 \cos \psi, R_1 \sin \psi)$ and the red cross indicates the position of the Kuramoto order parameter at the moment of the snapshot. Panels (d) and (e) are two consecutive snapshots, where the blue and purple color indicate which cluster the oscillator belong to.} 
	\label{fig.states}
\end{figure}

Before analyzing the model, we point out that \eqref{eq.vnpglophase} displays a wide variety of stable collective dynamics whose snapshots are shown in Fig.~\ref{fig.states}. Those dynamics are triggered by the addition of a higher harmonic to the Kuramoto-Sakaguchi model \cite{CP16,CP21,hmm93,KK01}. Since $\eta_3\propto\mu^2$, for the specific coupling \eqref{eq.coupling}, terms of order $\mu^2$ are needed to describe those new dynamics.

\begin{figure}
	\includegraphics[width=0.9\linewidth]{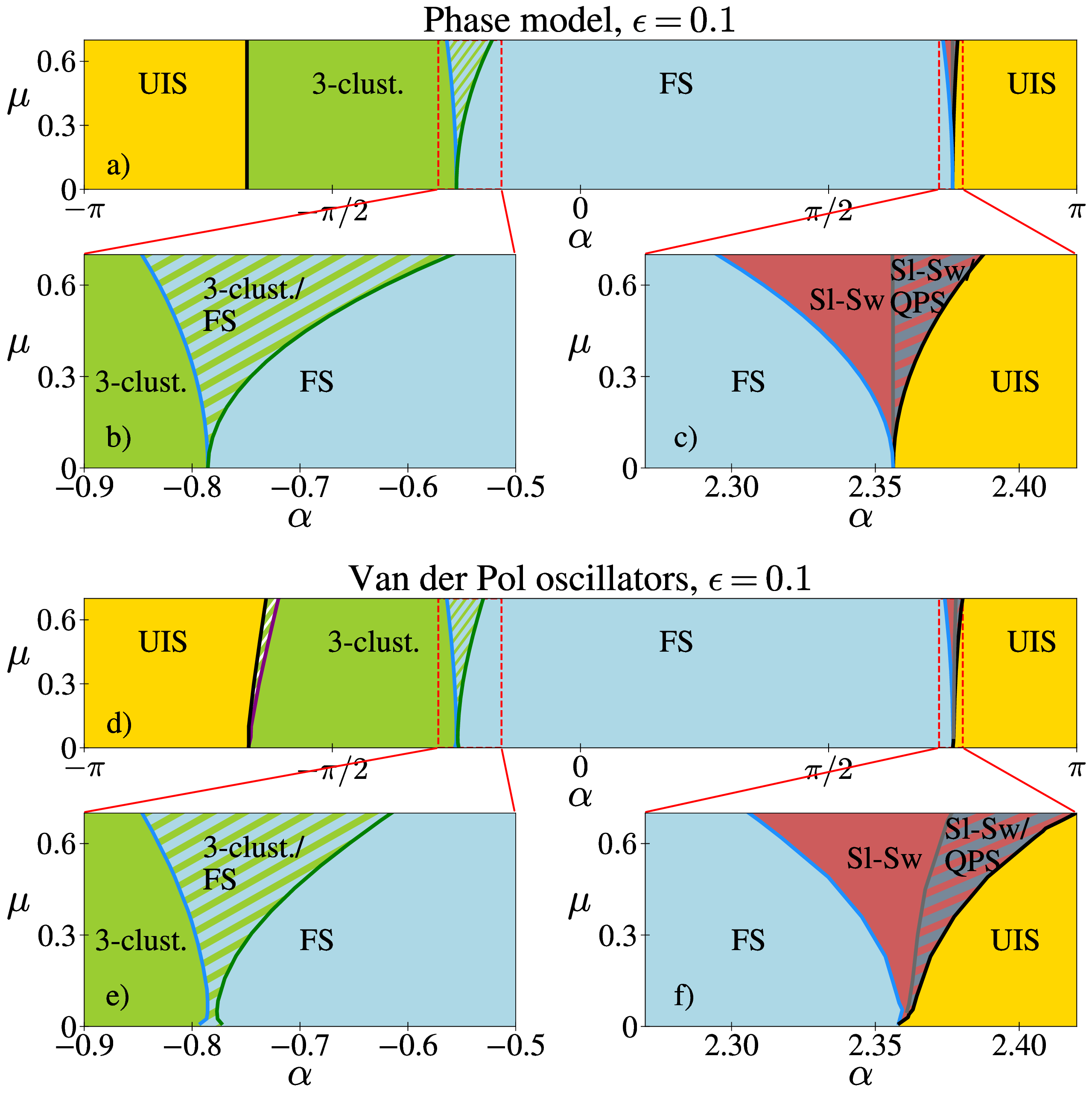}
	\caption
	{(a-c) Phase diagram predicted by the phase model (a) and magnified regions near the bifurcations [(b) and (c)]. In the yellow, blue, green, and red regions, UIS, FS, three cluster state, and slow switching are stable, respectively. In the hatched blue-green region, there is bistability between FS and three cluster state, while in the gray-red hatched region, both QPS and slow switching are stable. (d-f) Phase diagram obtained by numerical simulations of the Van der Pol oscillators (d) and magnified regions near the bifurcations [(e) and (f)]. The same color scheme as in panel (a-c) is used. The narrow green-white hatched region represents bistability between three-cluster state and other multi-cluster states. }
	\label{fig.globvdp}
\end{figure}

Similarly to the Kuramoto-Sakaguchi model, Eq.~\eqref{eq.vnpglophase} displays full synchrony (FS), where all oscillators form one point cluster, Fig.~\ref{fig.states}~(a), and uniform incoherence state (UIS), where oscillator phases are uniformly distributed, Fig.~\ref{fig.states}~(b). It can be proven that FS is stable when $\eta_{1s}+3\eta_{3}>0$ , expressed in the original parameters as:
	\begin{equation}
		\left(1-\frac{\mu^2}{8}\right)\sin\alpha+\left(1+\frac{\mu^2}{8}\right)\cos\alpha>0,
	\end{equation}
while UIS is stable when $\eta_{1s}$ and $\eta_3$ are negative:
	\begin{subequations}
		\begin{equation}
		\left(1+\frac{\mu^2}{64}\right)\sin\alpha+\left(1-\frac{7\mu^2}{64}\right)\cos\alpha<0,
		\end{equation}
		\begin{equation}
		-3\sin\alpha+5\cos\alpha<0.
		\end{equation}
	\end{subequations}
The regions where FS and UIS are stable are depicted in Fig.~\ref{fig.globvdp} in blue and yellow, respectively.

The phase model also allows us to analyze the stability of cluster states, in which oscillators form $n$ point clusters. The analysis indicates that the two-cluster states are unstable in the studied parameter region. 
%
%
Instead, we observe slow switching (Sl-Sw), where the two-cluster states are saddles, connected in a heteroclinic cycle~\cite{hmm93,KK01}.
As the system approaches the heteroclinic cycle, it switches between the saddle two-cluster states, Fig~\ref{fig.states}~(d). During the switches, one of the clusters is disintegrated and formed again, Fig~\ref{fig.states}~(e).
%
Slow switching is stable in the region where UIS and FS are unstable and $\eta_3<0$. This region of the parameter space is depicted in red in Fig.~\ref{fig.globvdp}.

In addition, it is possible to show that a three-cluster state characterized by $R_1=0$ and $R_3=1$, Fig.~\ref{fig.states}~(e), is stable when
	\begin{subequations}
	\begin{equation}
		\left(1+\frac{19\mu^2}{64}\right)\sin\alpha+\left(1-\frac{37\mu^2}{64}\right)\cos\alpha<0,
	\end{equation}
and
	\begin{equation}
		-3\sin\alpha+5\cos\alpha>0,
	\end{equation}
\end{subequations}
which is depicted with a green region in Fig.~\ref{fig.globvdp}.

The last state present in the phase model is quasi-periodic partial synchrony (QPS), Fig.~\ref{fig.states}~(c). In this state, the Kuramoto-Daido order parameters rotate uniformly, while individual oscillators evolve in a quasi-periodic fashion \cite{RP15,PR15,CP16}. This state arises when UIS becomes unstable and loses its stability via a subcritical Hopf bifurcation. Although the stability of QPS is not determined analytically, it can be computed without numerical simulations. We introduce an appropriate rotating frame of reference such that QPS is a fixed point and use the Newton-Raphson algorithm to find the fixed point and frequency of the rotating frame of reference. Then, a linear stability analysis of the fixed point is used to determine the stability of QPS. The region with stable QPS is depicted in Fig.~\ref{fig.globvdp} with gray hatching.

We sum up the previous discussions in Fig.~\ref{fig.globvdp} (a), where we depict the phase diagram of Eq.~\eqref{eq.vnpglophase} for $\epsilon=0.1$. Panels (b) and (c) depict magnifications of the regions close to bifurcations in Fig.~\ref{fig.globvdp} (a). The regions where FS, UIS, QPS, slow switching, and three-cluster state are stable are depicted in blue, yellow, gray, red, and green, respectively, and the hatching indicates bistability between two of those states. We stress that all the stability boundaries but the one for QPS are analytically determined. See \ref{app.analy} for the analytical derivations. We would like to remark that the nonlinearity of the Van der Pol oscillator is responsible for the complex behavior observed in the phase diagram, mainly due to the effect of the third- harmonic component.
See also e.g.,~\cite{Gol11} for the effect of high-harmonic components on the dynamics of the Van der Pol oscillator.

To check the accuracy of the results obtained from the phase model, we perform numerical simulations of an ensemble of $N=1000$ Van der Pol oscillators for $\epsilon=0.1$. In Fig.~\ref{fig.globvdp} (d), we depict the phase diagram obtained from numerical simulations for the ensemble of Van der Pol oscillators, with some magnified regions in panels (e) and (f), and the same color code as Fig.~\ref{fig.globvdp} (a). It is remarkable that almost all the dynamical states predicted analytically by the phase models are realized in the original system of Van der Pol oscillators. The similarities between the phase diagrams for the Van der Pol oscillators and the reduced phase  model prove the latter is capturing the qualitative dynamics with great accuracy.

There is only a small region in the parameter space [white-green hatched in Fig.~\ref{fig.globvdp}(d)] where there is bistability between the three-cluster state and a multi-cluster state not present in the phase model. 
The fact that this region shrinks as $\epsilon\rightarrow0$ indicates that higher-order terms in $\epsilon$ are needed to capture those dynamics. Additionally, we observe some discrepancy for small $\mu$. This discrepancy is not surprising; it is due to the limit cycle being less attractive as $\mu\rightarrow0$, and thus the assumption $\epsilon\ll\lambda/\omega$ for the phase reduction does not hold, given the fixed $\epsilon=0.1$. 


\section{Conclusions}

Through this work, we have presented a method to analytically obtain the frequency, limit cycle, and phase sensitivity function as a perturbative expansion in $\mu$ for a wide class of oscillators. These results enabled us to analytically derive reduced phase models for perturbed or coupled oscillators, facilitating an analytical understanding of their dynamics. We have exemplified the methodology with the Van der Pol oscillator, showing its usefulness to describe the dynamics and to obtain analytical phase diagrams.

We point to the importance of these results that broaden analytical phase reduction to a wider variety of systems. This is especially remarkable for the Van der Pol oscillator since it has multiple applications and is commonly used as a test-bed model \cite{PRK01,Guckenheimer,GM88}.

A possible extension of this work is to consider heterogeneous oscillators, in which each unit has a different frequency. The presented methodology is easily applicable to this case, though we have restricted ourselves to the identical case for the sake of simplicity.  We stress that the present approach can also yield phase models with higher-order interactions~\cite{battiston20} if the system contains multi-body interactions. Finally, extending the present approach to 
compute $\epsilon^2$ terms, where higher-order interactions appear naturally~\cite{leon19,gengel21,leon22a}, is also possible, though it is more challenging and falls beyond the scope of this work.

All in all, in this paper, we presented a systematic perturbative method to derive analytical expressions for the phase model, which broadens the knowledge about phase reduction. We believe the advances made in phase reduction are an indispensable step to understand the ubiquitous oscillatory dynamics. 

\section*{Acknowledgment}
We acknowledge JSPS KAKENHI JP22K11919, JP22H00516, and JST CREST JP-MJCR1913 for financial support. We thank Diego Paz\'o for the fruitful discussions and proposing the perturbation of Stuart-Landau oscillator with constant bias. 

\newpage
\appendix
\section{Deriving frequency, limit cycle, and phase sensitivity function of the Van der Pol oscillator}\label{app.VP}


In this section, we explicitly derive the frequency, limit cycle, and phase sensitivity function of the Van der Pol oscillator \cite{VdP26,VdP27}, following the method explained in the main text. We consider the evolution of the Van der Pol oscillator as written in the form of \eqref{eq.vandp} where $\mu$ is a small parameter. 

\subsection{Frequency and limit cycle}

In order to obtain the frequency $\omega$ and the limit cycle ${\boldsymbol X}_c = (x_c, y_c)$, we employ the Poincar\'e-Lindstedt method \cite{Nayf93,ABF18}. This method only seeks the periodic solution of \eqref{eq.vandp}, and any transient dynamic before the system reaches the limit cycle will be ignored. The first step of the method is making the change of variable $\theta=\omega t$, with unknown $\omega$. Then, we expand in powers of $\mu$ the frequency $\omega=\omega_0+\mu\omega_1+\dots$, and the limit-cycle solution $x_c=x_0+\mu x_1+\dots$ and $y_c=y_0+\mu y_1+\dots$. Plugging those expansions into Eq.~\eqref{eq.vandp} and collecting each order of $\mu$, we obtain an ordinary differential equation (ODE) for each order.

The $O(1)$ ODE yields the harmonic oscillator:
\begin{eqnarray} \label{eq.vdpo0}
	\omega_0\frac{dx_0}{d\theta}&=&y_0,\nonumber\\
	\omega_0\frac{dy_0}{d\theta}&=&-x_0,
\end{eqnarray}
where $\omega_0 =-1$\footnote{One can alternatively consider $\omega_0=1$, which is equivalent to defining the phase when $\mu=0$ as $\theta=\arctan(-y_0/x_0)$. In order to obtain the expressions 
for this choice of the sign of $\omega_0$, it is only necessary to change $\omega\rightarrow-\omega$, $(x_c(\theta),y_c(\theta))\rightarrow(x_c(-\theta),y_c(-\theta))$, and $\boldsymbol{Z}(\theta)\rightarrow-\boldsymbol{Z}(-\theta)$.} is determined from \eqref{eq.vandp} with $\mu=0$ . We consider an initial condition on the limit cycle, $y=0$ and $x = A_0+\mu A_1+\mu^2 A_2 \dots >0$, where the constants $A_n$ have to be determined later. The solution to \eqref{eq.vdpo0} is then: 
%
\begin{equation}\label{eq.vdpsolo0}
	x_0(\theta)=A_0\cos\theta;\quad y_0(\theta)=A_0\sin\theta.
\end{equation}
This solution is not yet determined due to the presence of $A_0$. We compute the next order to determine $A_0$.

The order $\mu$ ODE is given by:
\begin{eqnarray} \label{eq.vdpo1}
	\omega_0\frac{dx_1}{d\theta}&=&y_1-\omega_1\frac{dx_0}{d\theta},\nonumber\\
	\omega_0\frac{dy_1}{d\theta}&=&-x_1+(1-x_0^2)y_0-\omega_1\frac{dy_0}{d\theta}.
\end{eqnarray}
After plugging Eq.~\eqref{eq.vdpsolo0} into Eq.~\eqref{eq.vdpo1}, we find the solution to this linear nonhomogeneous system as the homogeneous solution (harmonic oscillations) plus a particular solution. The latter is easy to find since all terms in the r.h.s. are combinations of sines and cosines. This leads us to the solution to Eq.~\eqref{eq.vdpo1}:
\begin{subequations}\label{eq.vdpsolo1}
	\begin{multline}
		x_1(\theta)=\left(A_1+\theta\frac{A_0^3 -4 A_0}{8}  \right)\cos\theta\\
		+\left(\frac{A_0}{2}-\frac{7}{32}A_0^3-A_0 \theta  \omega_1\right) \sin \theta+\frac{1}{32} A_0^3 \sin 3 \theta,
	\end{multline}
	\begin{multline}
		y_1(\theta)=\left(A_1+\theta\frac{A_0^3-4A_0}{8}\right)\sin \theta\\
		+\frac{1}{32} \left(3 A_0^3+32 A_0 \theta \omega_1\right)\cos \theta-\frac{1}{32} 3 A_0^3 \cos 3 \theta.
	\end{multline}
\end{subequations}
We observe that all the secular terms, for example $\theta\frac{A_0^3 -4 A_0}{8}\cos\theta$ or $\theta\omega_1A_0\sin\theta$, are removed by choosing $A_0=2$ and $\omega_1=0$. Thus, the $O(1)$ solution, Eq.~\eqref{eq.vdpsolo0}, is completely determined. The $O(\mu)$ solution, Eq.~\eqref{eq.vdpsolo1}, is now free of the secular terms, but it is still not completely determined due to the presence of the constant $A_1$.
We can proceed to the following orders to determine $A_1$ and to obtain the frequency and limit cycle as a power expansion in $\mu$ to any desired order.
%
%
%
Following this procedure, we compute the frequency and limit-cycle solution up to order $\mu^2$,
%
%
\begin{equation}
	\omega=-1+\frac{\mu^2}{16}+O(\mu^4),
\end{equation}
\begin{multline}
	x_c(\theta)=2\cos\theta-\frac{\mu}{4}\left(3\sin\theta-\sin3\theta\right)\\
	-\frac{\mu^2}{8}\left(\cos \theta-\frac{3}{2} \cos3\theta+\frac{5}{12} \cos5\theta\right)+O(\mu^3),
\end{multline}
\begin{multline}
	y_c(\theta)=2\sin\theta+\frac{3\mu}{4}\left(\cos\theta-\cos3\theta\right)\\
	-\frac{\mu^2}{8}\left(2\sin \theta-\frac{9}{2} \sin3\theta+\frac{25}{12} x\sin5\theta\right)+O(\mu^3),
\end{multline}
which represents the limit cycle ${\boldsymbol X}_c$ of Eq.~\eqref{eq.vandp}.
%
%
Two details that worth mentioning: first, the frequency $\omega$ only contains even powers of $\mu$; and second, each new order in the limit-cycle solution adds a new harmonic, and correct the coefficients of the previous orders.

\subsection{Phase sensitivity function}
Now that the limit cycle ${\boldsymbol X}_c$ and frequency $\omega$ have been computed, we can use those expressions to obtain the phase sensitivity function ${\boldsymbol Z}$ as a power expansion in $\mu$. We extend the Poincar\'e-Lindstedt method to the adjoint equation for ${\boldsymbol Z}$:
\begin{equation}\label{eq.adjA}
	\omega\frac{d \boldsymbol{Z}(\theta)}{d\theta}=-J^\top(\boldsymbol{X}_c(\theta)) \boldsymbol{Z}(\theta)
\end{equation}
with the normalization condition 
\begin{equation}\label{eq.ncond}
	\boldsymbol{Z}(\theta)\cdot\boldsymbol{F}(\theta)=\omega.
\end{equation}

We expand the phase sensitivity function, $\boldsymbol{Z}=(Z_x,Z_y)$, in powers of $\mu$:
\begin{equation}
	Z_x=Z_{x0}+\mu Z_{x1}+\mu^2 Z_{x2}+\dots;\quad Z_y=Z_{y0}+\mu Z_{y1}+\mu^2 Z_{y2}+\dots
\end{equation}
It is important to have in mind that since $\boldsymbol{Z}$ is $2\pi$-periodic, all terms in the expansion are also $2\pi$-periodic, and thus, they do not contain secular terms. In other words, we seek for the solution to the adjoint equation in the space of $2\pi$-periodic functions.
We plug this expansion and the expressions for the frequency and the limit cycle into the adjoint equation, Eq.~\eqref{eq.adjA}, and normalization condition, Eq.~\eqref{eq.ncond}, and collect the terms at each order of $\mu$. This provides an ODE and a normalization condition for each order.

The $O(1)$ yields:
\begin{eqnarray}
	\omega_0\frac{dZ_{x0}}{d\theta}&=&Z_{y0},\nonumber\\
	\omega_0\frac{dZ_{y0}}{d\theta}&=&-Z_{x0},
\end{eqnarray}
with the normalization condition
\begin{equation}
	y_0 Z_{x0}-x_0 Z_{y0}=\omega_0.
\end{equation}
The solution to this system is
\begin{subequations}\label{eq.vdpZo0}
	
	\begin{equation}
		Z_{x0}(\theta)=B_0\cos\theta-\frac{1}{2}\sin\theta,
	\end{equation}
	\begin{equation}
		Z_{y0}(\theta)=B_0\sin\theta+\frac{1}{2}\cos\theta,
	\end{equation}
\end{subequations}
where $B_0$ is an integration constant, which is not determined yet. We compute the next order to determine $B_0$.

The ODE and normalization condition for order $\mu$ are:
\begin{eqnarray}
	\omega_0\frac{dZ_{x1}}{d\theta}&=&Z_{y0}+2x_0y_0Z_{y0}-\omega_1\frac{dZ_{x0}}{d\theta},\nonumber\\
	\omega_0\frac{dZ_{y1}}{d\theta}&=&-Z_{x0}-(1-x_0^2)Z_{y0}-\omega_1\frac{dZ_{y0}}{d\theta},
\end{eqnarray}
and
\begin{equation}
	y_0 Z_{x1}-x_0 Z_{y1}+y_1Z_{x0}-x_1 Z_{y0}+(1-x_0^2)y_0Z_{y0}=\omega_1,
\end{equation}
whose solution is computed as a sum of the homogeneous solution and a particular solution as
\begin{subequations}\label{eq.vdpZo1}
	\begin{multline}
		Z_{x1}(\theta)= \left(-B_0 \theta +B_1+\frac{9}{8}\right)\cos \theta\\
		-\frac{7}{8} B_0 \sin \theta+\frac{5}{8} B_0 \sin 3\theta+\frac{5}{16} \cos 3\theta,
	\end{multline}
	\begin{multline}
		Z_{y1}(\theta)=\left(-B_0 \theta +B_1+\frac{1}{8}\right)\sin \theta-\frac{1}{8} B_0 \cos \theta\\
		+\frac{1}{8} B_0 \cos 3\theta-\frac{1}{16}\sin3 \theta,
	\end{multline}
\end{subequations}
where $B_1$ is an undetermined integration constant. 
Since ${\boldsymbol Z}$ should be $2\pi$-periodic, we need to eliminate the secular terms from these equations.
It is easy to spot that, if $B_0=0$, all the secular terms are removed. Proceeding forward, we can determine the phase sensitivity function to any desired order. The result up to order $\mu^2$ is:
\begin{multline}
	Z_x(\theta)=-\frac{1}{2}\sin\theta+\frac{\mu}{16}\left(15\cos\theta+5\cos3\theta\right)\\
	-\frac{\mu^2}{64}\left(21\sin\theta+\sin3\theta-\frac{29}{6}\sin5\theta\right)+O(\mu^3),
\end{multline}
\begin{multline}
	Z_y(\theta)=\frac{1}{2}\cos\theta-\frac{\mu}{16}\left(\sin\theta+\sin3\theta\right)\\
	+\frac{\mu^2}{64}\left(
	3\cos\theta-5\cos3\theta-\frac{1}{6}\cos5\theta\right)+O(\mu^3).
\end{multline}

As in the expression for the limit cycle, higher-order corrections add new harmonics and correct the coefficients in the previous orders.

\section{Stuart-Landau oscillator with a constant bias} \label{app.SLcb}

In this section, to demonstrate the generality of our approach, we consider as an example a  Stuart-Landau oscillator with a constant bias, 
$\dot{A}=(1+i\nu)A-|A|^2A+\mu$, written in polar coordinates $A=re^{i\phi}$ as
\begin{eqnarray}
	\dot{r}&=&r(1-r^2)+\mu\cos\phi,\\
	\dot{\phi}&=&\nu-\frac{\mu}{r}\sin\phi,
\end{eqnarray}
where $\nu$ is a free parameter. Following the procedure explained above, we obtain the expressions for the frequency, limit cycle, and phase sensitivity function up to order $\mu^2$:
\begin{equation}
	\omega=\nu-\mu^2\frac{2\mu^2}{\nu(4+\nu^2)}+O(\mu^4),
\end{equation}
\begin{subequations}
	\begin{multline}
		r=1+\mu\frac{2 \cos\theta+\nu  \sin\theta}{\nu ^2+4}-\frac{\mu^2}{\nu ^2+4}\bigg[\frac{3}{4}+\cos\theta-\frac{2 \sin \theta}{\nu }\\
		-\frac{\left(\nu ^4+23 \nu ^2+4\right) \cos 2 \theta}{4 \left(\nu ^2+1\right) \left(\nu ^2+4\right)}+\frac{\left(-\nu ^4+16 \nu ^2+8\right) \sin 2\theta}{2 \nu\left(\nu ^2+4\right) \left(\nu ^2+1 \right)}\bigg],
	\end{multline}
	\begin{multline}
		\phi=\theta+\mu\frac{\cos\theta-1}{\nu}+\mu^2\bigg[\frac{1-\cos 2\theta}{2 \nu  \left(\nu ^2+4\right)}+\frac{\sin \theta}{\nu ^2}\\
		-\frac{\left(2+\nu ^2\right) \sin 2\theta}{2\nu ^2 \left(\nu ^2+4\right)}\bigg],
	\end{multline}
\end{subequations}
\begin{subequations}
	\begin{multline}
		Z_r=\mu\frac{\nu  \cos \theta+2 \sin \theta}{\nu ^2+4}+\frac{\mu^2}{\nu ^2+4}\bigg[\frac{8-4 \nu ^2}{\nu ^3+4 \nu }-\frac{2\cos\theta}{\nu}\\
		-\frac{2\nu\cos2\theta}{1+\nu^2}+\sin\theta-\frac{(5+\nu^2)\sin2\theta}{2(\nu+\nu^2)}\bigg],
	\end{multline}
	\begin{equation}
		Z_\phi=1+\mu\frac{\sin \theta}{\nu }+\frac{\mu^2}{(\nu^2+4)}\bigg[\frac{8+6 \nu ^2}{\nu ^2 \left(\nu ^2+4\right)}-\frac{2}{\nu^2}\cos\theta-\frac{\sin2\theta}{\nu}\bigg].
	\end{equation}
\end{subequations}	

\begin{figure}
	\includegraphics[width=\linewidth]{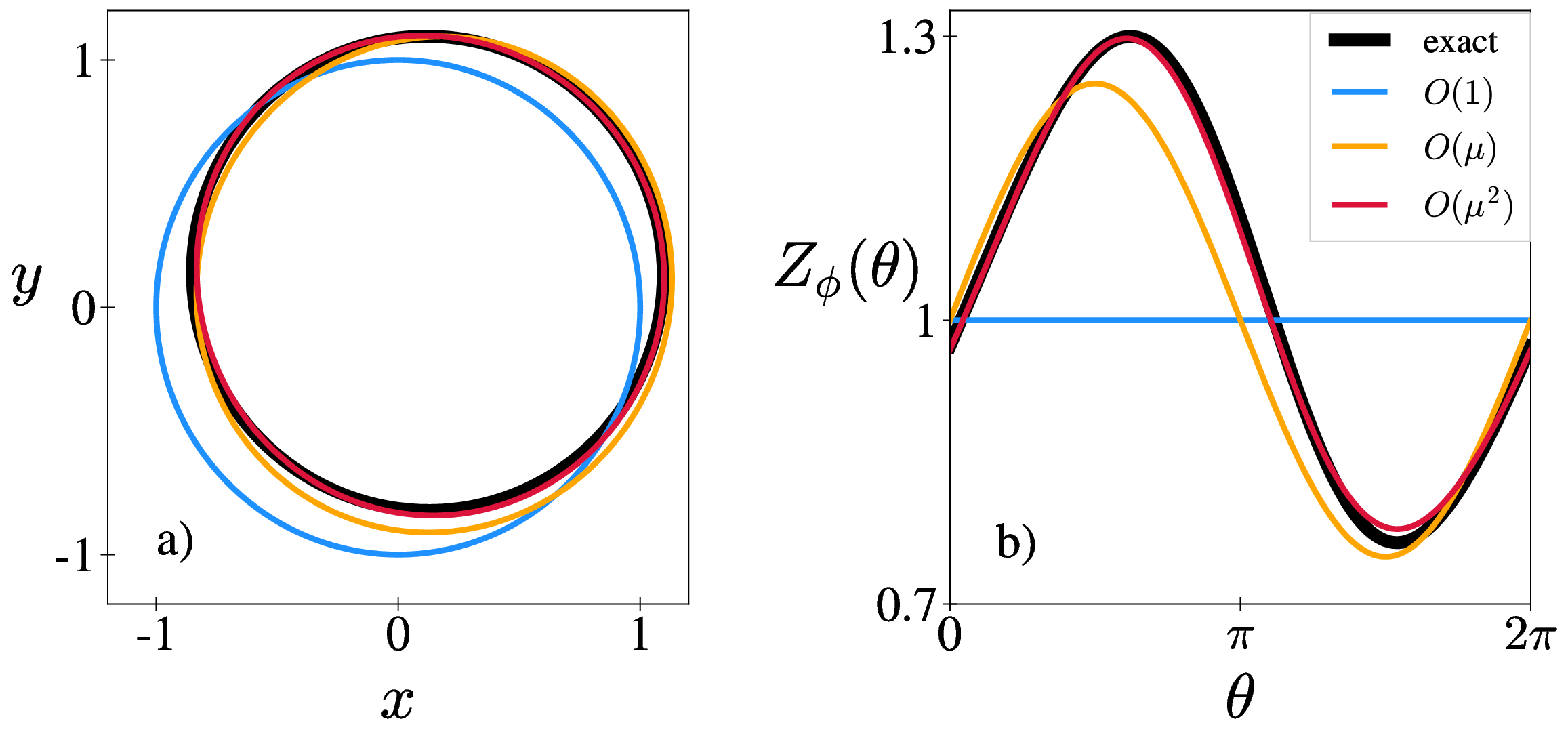}
	\caption
	{Limit cycle of the Stuart-Landau oscillator with a constant bias for $\mu=0.5$ and $\nu=2$ (a) and the angular component of its phase sensitivity function. The exact numerical limit cycle and the $O(1)$, $O(\mu)$ and $O(\mu^2)$ approximations are depicted in black, blue, orange, and red, respectively.}
	\label{fig.SL}
\end{figure}

In figure~\ref{fig.SL}~(a), we compare the approximate limit cycles of the Stuart-Landau with a constant bias to the exact one obtained by direct numerical simulation for $\mu=0.5$ and $\nu=2$. Although the exact limit cycle (black) is close to be circular, it is far from the unperturbed Stuart-Landau limit cycle (blue), i.e., the $O(1)$ approximation. The higher-order corrections contain new harmonics that reduce this discrepancy, as can be seen in the $O(\mu)$ (orange) and $O(\mu^2)$ (red) approximations. In panel (b) we depict the angular component of the phase sensitivity function for the exact numerical simulations in black and the $O(1)$, $O(\mu)$ and $O(\mu^2)$ approximations in blue, orange, and red, respectively. The conclusions about accuracy are equivalent to the ones obtained for the Van der Pol oscillator.

\section{Analyzing the phase model Eq.~\eqref{eq.vnpglophase} }\label{app.analy}

In this section, we derive the analytical stability conditions of the dynamical regimes of the reduced phase model for the ensemble of globally coupled Van der Pol oscillators, Eq.~\eqref{eq.vnpglophase}. Without loss of generality, we bring the phase model to the form $\dot{\theta}_j=\sum_{k}\Gamma(\theta_k-\theta_j)/N$, with
\begin{equation}\label{eq.SMphmodel}
	\Gamma(x)=\epsilon\bigg[\eta_{1s} \sin x+\eta_{1c} \cos x
	+\eta_3 \sin (3x)\bigg],
\end{equation}
by choosing a rotating reference frame with the frequency $\Omega$ of the collective oscillation.

Although the stability analysis of the states can be performed for a finite number of oscillators, we consider the thermodynamic limit $N\rightarrow\infty$ for simplicity. In this limit, we can define a probability density function of the oscillators $\rho(\theta,t)$, where {$\rho(\theta, t) d\theta$} measures the fraction of oscillators between $\theta$ and $\theta+d\theta$ at time $t$. This probability density evolves following the continuity equation $\partial_t\rho=-\partial_\theta(\rho\dot{\theta})$. We remark that the Kuramoto-Daido order parameters are defined in this limit as $R_ne^{i\psi_n}=\int_0^{2\pi} \rho e^{in\theta} d\theta$ for $n=1, 2, \ldots$.

\subsection{Full synchronization}
We begin our analysis {with the study of} full synchronization (FS). Full synchronization is the state in which all oscillators form one point cluster, and thus $R_1=R_3=1$. It is known that for coupling functions of the form \eqref{eq.SMphmodel}, the linear stability analysis of FS yields a zero exponent, associated with a global phase shift, plus the exponents associated with variations of individual oscillators. These perturbations grow with an exponent
\begin{equation}
	\lambda_{FS}=-\eta_{1s}-3\eta_3.
\end{equation}
Thus, when $\eta_{1s}+3\eta_3>0$, FS is stable as stated in the main text.

\subsection{Uniform incoherent state}\label{sec.UIS}
The uniform incoherent state (UIS) is characterized by vanishing order parameters $R_1=R_3=0$, and thus the oscillator phases are uniformly distributed. It is much easier to compute the stability of UIS if we decompose the probability density of the oscillators, $\rho$, into its Fourier modes:
\begin{equation}
	\rho(\theta,t)=\frac{1}{2\pi}\sum_{k=-\infty}^{\infty}\rho_k(t)e^{i k \theta},
\end{equation}
where $\rho_{-k}=\rho_k^*$ and $\rho_0=1$. Notice that the coefficients $\rho_k$ correspond to Kuramoto-Daido order parameters $R_ke^{i\psi_k}=\int{_0^{2\pi}} \rho e^{ik\theta} {d\theta}=\rho_{-k}$. Plugging this decomposition into the continuity equation, we obtain the evolution of each mode $\rho_k$:
\begin{multline}\label{eq.fourmodes}
	\dot{\rho}_k=\frac{\epsilon k}{2}\bigg[(\eta_{1s}-i\eta_{1c})\rho_1\rho_{k-1}-(\eta_{1s}+i\eta_{1c})\rho_{-1}\rho_{k+1}\\
	+\eta_3\big(\rho_3\rho_{k-3}-\rho_{-3}\rho_{k+3}\big)\bigg].
\end{multline}
In this representation, the UIS is characterized by $\rho_{n\not=0}=0$. The linear stability analysis of UIS is now straightforward to perform. It indicates that UIS is stable when $\eta_{1s}$ and $\eta_{3}$ are negative, as stated in the main text.

\subsection{Quasiperiodic partial synchrony}
In this section, we indicate how the analysis of QPS is performed. As in the analysis of UIS, it is convenient to define a probability density $\rho$ of the oscillators and work in the Fourier modes. 
Then, QPS is characterized by the density of the oscillators rotating uniformly with frequency $\tilde{\Omega}$, 
which means that the Fourier mode $\rho_k$ of $\rho$ evolves as
	\begin{equation}
		\rho_k=e^{ik\tilde{\Omega}t}\tilde{\rho}_k,
	\end{equation}
where $\tilde{\rho}_k$ is a constant.
In the rotating frame of reference with frequency $\tilde{\Omega}$, QPS corresponds to a continuous family of fixed points of equation \eqref{eq.fourmodes}, where 
all 
the fixed points are related through a rotation. We can select one of the fixed points, for example, by imposing $\operatorname{Re}(\tilde{\rho}_1)=0$, and determine the values of $\tilde{\rho}_k$ and frequency $\tilde{\Omega}$ using the Newton-Raphson algorithm. Once those values are determined, we perform linear stability analysis of the state. The spectrum will contain one zero exponent related to the rotational symmetry of QPS and the other exponents determine the stability of QPS. Because \eqref{eq.fourmodes} is an infinite set of ODEs, it is is necessary to truncate the system numerically, keeping only a finite number of modes.

\subsection{Two-cluster states}
The two-cluster state is characterized by a fraction $p$ of the oscillators being in a point cluster $A$ with phase $\theta_A$, while the other $(1-p)$ fraction of the oscillators form cluster $B$ with phase $\theta_B$. Since all oscillators are identical, to completely specify a two-cluster state, it is enough to know the phase difference $\Delta=\theta_A-\theta_B$ and the fraction $p$ of the oscillators in cluster $A$.

The evolution of the phase of each cluster is given by:
\begin{eqnarray}
	\dot{\theta}_A&=&p\Gamma(0)+(1-p)\Gamma(\theta_B-\theta_A),\\	\dot{\theta}_B&=&(1-p)\Gamma(0)+p\Gamma(\theta_A-\theta_B),
\end{eqnarray}
and thus the phase difference $\Delta$ evolves according to:
\begin{equation}\label{eq.delta}
	\dot{\Delta}=(2p-1)\Gamma(0)+(1-p)\Gamma(-\Delta)-p\Gamma(\Delta).
\end{equation}
The fixed points of this equation are the possible two-cluster states, and thus the relation between $p$ and $\Delta$ is:
\begin{equation}
	p(\Delta)=\frac{\Gamma(0)-\Gamma(-\Delta)}{2\Gamma(0)-\Gamma(\Delta)-\Gamma(-\Delta)}.
\end{equation}
Any $\Delta$ such that $p(\Delta)\in(0,1)$ gives a possible two-cluster state, nevertheless, it is not necessarily stable. To analyze the stability, we use the fact that any perturbation given to the state can be decomposed into three orthogonal modes \cite{GHS+92,KK01}. One mode corresponds to the locking of the two clusters and the other two represent the disintegration of the individual clusters. We denote their corresponding eigenvalues as $\lambda_L$, $\lambda_A$, and $\lambda_B$. 

We first analyze the locking of the two clusters by studying a perturbation given to $\Delta$. Linearizing \eqref{eq.delta} around the fixed point $\Delta^*$, we observe that the perturbation grows with the associated exponent
\begin{equation}
	\lambda_L=-(1-p)\Gamma'(-\Delta^*)-p\Gamma'(\Delta^*).
\end{equation}
The disintegration of cluster $A$ is analyzed by computing the effect of perturbing one oscillator from cluster $A$. This perturbation will grow with the exponent
\begin{equation}
	\lambda_A=-p\Gamma'(0)-(1-p)\Gamma'(-\Delta^*).
\end{equation}
The exponent associated with the disintegration of cluster $B$ is obtained by changing $p\rightarrow(1-p)$ and $\Delta^*\rightarrow-\Delta^*$ in the previous equation. For the coupling function Eq.~\eqref{eq.SMphmodel}, at least one of the eigenvalues is positive in the studied parameter space, hence two-cluster states are always unstable in the studied parameter region.

\subsection{Slow switching}
A dynamical regime closely related to the two-cluster states is the slow switching, a state in which two unstable two-cluster states are connected forming an attractive heteroclinic cycle. It has been shown that for this state to be stable, the following conditions have to be satisfied \cite{KK01}:
\begin{itemize}
	\item A value of $p$ exists such that there are three unstable two-cluster states with phase differences $0<\Delta_{1}<\Delta_2<\Delta_{3}<2\pi$. We denote their eigenvalues by $\lambda_{L,A,B}^{n}$.
	\item Full synchrony is unstable.
	\item $\lambda_L^{2}>0$ while $\lambda_L^{1}<0$ and $\lambda_L^{3}<0$.
	\item $\lambda_A^{1}>0$ and $\lambda_B^{1}<0$ while $\lambda_A^{3}<0$ and $\lambda_B^{3}>0$.
\end{itemize}
The coupling function Eq.~\eqref{eq.SMphmodel} verify those conditions in the parameter space where UIS and FS are unstable and $\eta_3<0$, as stated in the main text.

\subsection{Three-cluster states}
The last state we analyze is the three-cluster state characterized by $R_1=0$ and $R_3=1$. The values of the Kuramoto-Daido order parameters imply that each cluster is formed by one third of the oscillators and the phase difference between consecutive clusters is $2\pi/3$. As was done for the two-cluster state, we can decompose the perturbation into different modes, i.e., the locking of the clusters and their disintegration. 

The locking of the clusters is analyzed by computing the evolution of the phase differences between the clusters, $\Delta_1=\theta_A-\theta_B$ and $\Delta_2=\theta_B-\theta_C$. The phase differences evolve as:
\begin{eqnarray}
	3\dot{\Delta}_1&=&\Gamma(-\Delta_1)-\Gamma(\Delta_1)+\Gamma(-\Delta_1-\Delta_2)-\Gamma(-\Delta_2),\\
	3\dot{\Delta}_2&=&\Gamma(\Delta_1)-\Gamma(\Delta_2)-\Gamma(\Delta_1+\Delta_2)+\Gamma(-\Delta_2).
\end{eqnarray}
The linear stability analysis of the above equations indicates that the perturbation applied to the fixed point $\Delta_1=\Delta_2=2\pi/3$ grows with exponents: 
\begin{equation}
	\lambda_1^{\pm}=\frac{3}{2} (\eta_{1s}-6 \eta_3\pm i \eta_{1c}),
\end{equation}
which are mutually complex conjugate.

To analyze the disintegration of one cluster, we follow the same procedure as in full synchrony but with $R_1=0$. This yields the exponent
\begin{equation}
	\lambda_2=-\eta_3.
\end{equation}
Thus, the three-cluster state is stable when $\eta_{1s}-6 \eta_3<0$ and $\eta_3>0$, yielding the expressions Eq.~(13) in the main text.


\begin{thebibliography}{10}

	\bibitem{PRK01}
	A.~S. Pikovsky, M.~G. Rosenblum, J.~Kurths, Synchronization, a Universal
	Concept in Nonlinear Sciences, Cambridge University Press, Cambridge, 2001.
	
	\bibitem{Win67}
	A.~T. Winfree, Biological rhythms and the behavior of populations of coupled
	oscillators., J. Theor. Biol. 16 (1967) 15--42.
	
	\bibitem{Win80}
	A.~T. Winfree, The Geometry of Biological Time, Springer, New York, 1980.
	
	\bibitem{Kur75}
	Y.~Kuramoto, Self-entrainment of a population of coupled non-linear
	oscillators, in: H.~Araki (Ed.), International Symposium on Mathematical
	Problems in Theoretical Physics, Vol.~39 of Lecture Notes in Physics,
	Springer, Berlin, 1975, pp. 420--422.
	
	\bibitem{Kur84}
	Y.~Kuramoto, Chemical Oscillations, Waves, and Turbulence, {S}pringer-{V}erlag,
	Berlin, 1984.
	
	\bibitem{Str00}
	S.~H. Strogatz, From {K}uramoto to {C}rawford: exploring the onset of
	synchronization in populations of coupled oscillators, Physica D 143 (2000)
	1--20.
	
	\bibitem{BGL+20}
	C.~Bick, M.~Goodfellow, C.~R. Laing, E.~A. Martens, Understanding the dynamics
	of biological and neural oscillator networks through exact mean-field
	reductions: a review, The Journal of Mathematical Neuroscience 10~(1) (2020)
	1--43.
	
	\bibitem{mertens11}
	D.~Mertens, R.~Weaver, Synchronization and stimulated emission in an array of
	mechanical phase oscillators on a resonant support, Phys. Rev. E 83 (2011)
	046221.
	
	\bibitem{EK86}
	B.~Ermentrout, N.~Kopell, Parabolic bursting in an excitable system coupled
	with a slow oscillation, {SIAM} J. Appl. Math. 46 (1986) 233--253.
	
	\bibitem{Gol10}
	D.~S. Goldobin, J.-n. Teramae, H.~Nakao, G.~B. Ermentrout,
	{Dynamics of
		limit-cycle oscillators subject to general noise}, Phys. Rev. Lett. 105
	(2010) 154101.
		
	\bibitem{Jun09}
	J.-n. Teramae, H.~Nakao, G.~B. Ermentrout,
	{Stochastic
		phase reduction for a general class of noisy limit cycle oscillators}, Phys.
	Rev. Lett. 102 (2009) 194102.
	
	
	\bibitem{leon20}
	I.~Le\'on, D.~Paz\'o,
	{Quasi phase
		reduction of all-to-all strongly coupled
		$\ensuremath{\lambda}\ensuremath{-}\ensuremath{\omega}$ oscillators near
		incoherent states}, Phys. Rev. E 102 (2020) 042203.

	\bibitem{Kure22}
	W.~Kurebayashi, T.~Yamamoto, S.~Shirasaka, H.~Nakao,
{Phase
		reduction of strongly coupled limit-cycle oscillators}, Phys. Rev. Res. 4
	(2022) 043176.

	\bibitem{leon19}
	I.~Le\'on, D.~Paz\'o,
{Phase reduction
		beyond the first order: The case of the mean-field complex
		{G}inzburg-{L}andau equation}, Phys. Rev. E 100 (2019) 012211.

	
	\bibitem{WilErm2019}
	D.~Wilson, B.~Ermentrout,
{Phase models
		beyond weak coupling}, Phys. Rev. Lett. 123 (2019) 164101.

	\bibitem{KuraNakao19}
	Y.~Kuramoto, H.~Nakao, On the concept of dynamical reduction: the case of
	coupled oscillators, Phil. Trans. R. Soc. A 377~(2160) (2019) 20190041.
	
	\bibitem{RP19}
	M.~Rosenblum, A.~Pikovsky, {Numerical
		phase reduction beyond the first order approximation}, Chaos 29~(1) (2019)
	011105.

	\bibitem{ZKN22}
	J.~Zhu, Y.~Kato, H.~Nakao,
{Phase
		dynamics of noise-induced coherent oscillations in excitable systems}, Phys.
	Rev. Res. 4 (2022) L022041.

	\bibitem{Win79}
	A.~T. Winfree, Twenty-four hard problems about the mathematics of 24-hour
	rhythms, in: Lectures in Applied Mathematics, Nonlinear Oscillations in
	Biology, Vol.~17, 1979, pp. 93--126.
	
	\bibitem{GM88}
	L.~Glass, M.~C. Mackey, {From
		Clocks to Chaos: The Rhythms of Life}, Princeton University Press, 1988.

	\bibitem{Nak16}
	H.~Nakao, {Phase reduction
		approach to synchronisation of nonlinear oscillators}, Contemporary Physics
	57~(2) (2016) 188--214.
	
	\bibitem{gengel21}
	E.~Gengel, E.~Teichmann, M.~Rosenblum, A.~Pikovsky,
{High-order phase reduction
		for coupled oscillators}, J. Phys.: Complexity 2~(1) (2021) 015005.
	
	\bibitem{leon22a}
	I.~Le\'on, D.~Paz\'o,
	{Enlarged
		{K}uramoto model: Secondary instability and transition to collective chaos},
	Phys. Rev. E 105 (2022) L042201.

	
	\bibitem{HI97}
	F.~C. Hoppensteadt, E.~M. Izhikevich, Weakly connected neural networks.,
	Spinger Verlag, N.Y., 1997.
	
	\bibitem{Izh07}
	E.~M. Izhikevich, Dynamical Systems in Neuroscience, The MIT Press, Cambridge,
	Massachusetts, 2007.
	
	\bibitem{ermentrout96}
	B.~Ermentrout, Type i membranes, phase resetting curves, and synchrony, Neural
	Comp. 8 (1996) 979--1001.
	
	\bibitem{BMH04}
	E.~Brown, J.~Moehlis, P.~Holmes,
	{{On the Phase Reduction and
			Response Dynamics of Neural Oscillator Populations}}, Neural Computation
	16~(4) (2004) 673--715.
		
	\bibitem{Nayf93}
	A.~H. Nayfeh, Introduction to Perturbation Techniques, Wiley, London, 1993.
	
	\bibitem{Strogatz}
	S.~H. Strogatz, Nonlinear Dynamics and Chaos, Addison-Wesley, Reading, 1995.
	
	\bibitem{ABF18}
	P.~Amore, J.~P. Boyd, F.~M. Fernández,
	{High order analysis of the limit
		cycle of the van der pol oscillator}, Journal of Mathematical Physics 59~(1)
	(2018) 012702.

	\bibitem{VdP26}
	B.~van~der Pol, Vii. forced oscillations in a circuit with non-linear
	resistance. (reception with reactive triode), The London, Edinburgh, and
	Dublin Philosophical Magazine and Journal of Science 3~(13) (1927) 65--80.

	
	\bibitem{VdP27}
	B.~Van~der Pol, J.~Van~der Mark, Frequency demultiplication, Nature 120 (1927)
	363--364.

	
	\bibitem{Dai96}
	H.~Daido, Onset of cooperative entrainment in limit-cycle oscillators with
	uniform all-to-all interactions: bifurcation of the order function, Physica D
	91 (1996) 24--66.
	
	\bibitem{CP16}
	P.~Clusella, A.~Politi, M.~Rosenblum,
{A minimal model of
		self-consistent partial synchrony}, New Journal of Physics 18~(9) (2016)
	093037.

	
	\bibitem{CP21}
	P.~Clusella, A.~Politi,
{Irregular collective
		dynamics in a kuramoto–daido system}, Journal of Physics: Complexity 2~(1)
	(2020) 014002.
	
	\bibitem{hmm93}
	D.~Hansel, G.~Mato, C.~Meunier,
{Clustering and slow
		switching in globally coupled phase oscillators}, Phys. Rev. E 48 (1993)
	3470--3477.
	
	\bibitem{KK01}
	H.~Kori, Y.~Kuramoto,
{Slow switching in
		globally coupled oscillators: robustness and occurrence through delayed
		coupling}, Phys. Rev. E 63 (2001) 046214.
	
	\bibitem{RP15}
	M.~Rosenblum, A.~Pikovsky, Two types of quasiperiodic partial synchrony in
	oscillator ensembles, Phys. Rev. E 92 (2015) 012919.

	
	\bibitem{PR15}
	A.~Politi, M.~Rosenblum, Equivalence of phase-oscillator and integrate-and-fire
	models, Phys. Rev. E 91 (2015) 042916.

	
	\bibitem{Gol11}
	D.~S. Goldobin,
	{Anharmonic
		resonances with recursive delay feedback}, Physics Letters A 375~(39) (2011)
	3410--3414.
	
	
	\bibitem{Guckenheimer}
	J.~{G}uckenheimer, P.~Holmes, Nonlinear {O}scillations, {D}ynamical {S}ystems,
	and {B}ifurcations of {V}ector {F}ields, {S}pringer-{V}erlag, New York, 1983.
	
	
	\bibitem{battiston20}
	F.~Battiston, G.~Cencetti, I.~Iacopini, V.~Latora, M.~Lucas, A.~Patania, J.-G.
	Young, G.~Petri,
{Networks
		beyond pairwise interactions: Structure and dynamics}, Phys. Rep. 874 (2020)
	1 -- 92.
	
	\bibitem{GHS+92}
	D.~Golomb, D.~Hansel, B.~Shraiman, H.~Sompolinsky,
{Clustering in
		globally coupled phase oscillators}, Phys. Rev. A 45 (1992) 3516--3530.

	
\end{thebibliography}

\end{document}